\documentclass[12pt]{iopart}%
\usepackage{iopams}
\usepackage{amsfonts}
\usepackage{amssymb}
\usepackage{graphicx}%

\def\beqra{\begin{eqnarray}} \def\eeqra{\end{eqnarray}}
\def\beqast{\begin{eqnarray*}} \def\eeqast{\end{eqnarray*}}
\def\beq{\begin{equation}}      \def\eeq{\end{equation}}
\def\be{\begin{enumerate}}   \def\ee{\end{enumerate}}

\def\gam{\gamma}

\def\si{\sigma}

\def\del{\delta}

\def\om{\omega}

\def\pa{\partial}

\def\ce{{\cal E}}

\def\cm{{\cal M}}

\def\raisenot{\raise .5mm\hbox{/}}
\def\nota{\ \hbox{{$a$}\kern-.49em\hbox{/}}}
\def\notA{\hbox{{$A$}\kern-.54em\hbox{\raisenot}}}
\def\notb{\ \hbox{{$b$}\kern-.47em\hbox{/}}}
\def\notB{\ \hbox{{$B$}\kern-.60em\hbox{\raisenot}}}
\def\notc{\ \hbox{{$c$}\kern-.45em\hbox{/}}}
\def\notd{\ \hbox{{$d$}\kern-.53em\hbox{/}}}
\def\notbd{\ \hbox{{$D$}\kern-.61em\hbox{\raisenot}}} 
\def\note{\ \hbox{{$e$}\kern-.47em\hbox{/}}}
\def\notk{\ \hbox{{$k$}\kern-.51em\hbox{/}}}
\def\notp{\ \hbox{{$p$}\kern-.43em\hbox{/}}}
\def\notq{\ \hbox{{$q$}\kern-.47em\hbox{/}}}
\def\notW{\ \hbox{{$W$}\kern-.75em\hbox{\raisenot}}}
\def\notz{\ \hbox{{$Z$}\kern-.61em\hbox{\raisenot}}}
\def\notpa{\hbox{{$\partial$}\kern-.54em\hbox{\raisenot}}}

\def\fo{\hbox{{1}\kern-.25em\hbox{l}}}  

\def\tr{{\rm Tr}}

\def\sigx{\sigma(x)}
\def\pix{\pi(x)}

\begin{document}
\title[Fermion Bag Solitons: Inverse Scattering Analysis]
{Fermion Bag Solitons in the Massive Gross-Neveu and 
Massive Nambu-Jona-Lasinio Models in $1+1$ 
Dimensions: Inverse Scattering Analysis}
\author{Joshua Feinberg$^{a,b,c}$ and Shlomi Hillel$^b$}

\begin{abstract}
Formation of fermion bag solitons is an important paradigm in the theory of 
hadron structure. We report here on our non-perturbative analysis of this 
phenomenon in the $1+1$ dimensional massive Gross-Neveu model, in the large 
$N$ limit. Our main result is that the extremal static bag configurations are 
reflectionless, as in the massless Gross-Neveu model. Explicit formulas for 
the profiles and masses of these solitons are presented. We also present a 
particular type of self-consistent reflectionless solitons which arise in the 
massive Nambu-Jona-Lasinio models, in the large-$N$ limit.
\end{abstract}

\address{ $^a$Physics Department, University of Haifa at Oranim, Tivon 36006, 
Israel~\\ 
$^b$Physics Department, Technion, Israel Institute of Technology, Haifa 32000,
Israel~\\
$^c${\em Talk delivered by JF.}}
\ead{joshua@physics.technion.ac.il}

\section{Introduction}

An important dynamical mechanism, by which fundamental particles
acquire masses, is through interactions with vacuum condensates. Thus, 
a massive particle may carve out around itself a spherical region 
\cite{sphericalbag} or a shell \cite{shellbag} in which the
condensate is suppressed, thus reducing the effective mass of the particle
at the expense of volume and gradient energy associated with the
condensate. This picture has interesting phenomenological consequences
\cite{sphericalbag,mackenzie}.

This dynamical distortion of the homogeneous vacuum condensate configuration,
namely, formation of fermion bag solitons, was demonstrated explicitly 
by Dashen, Hasslacher and Neveu (DHN) \cite{dhn} many years ago, who studied 
semiclassical bound states in the $1+1$ dimensional Gross-Neveu 
(GN) model \cite{gn}, using the inverse scattering method \cite{inverse}. 
Following DHN, Shei \cite{shei} has applied the inverse scattering method to 
study solitons in the $1+1$ dimensional Nambu-Jona-Lasinio (NJL) model 
\cite{njl} in the large-$N$ limit. 

Fermion bags in the GN model were discussed in the 
literature several other times since the work of DHN, 
using alternative methods \cite{others, papa, josh1}. For a review on 
these and related matters (with an emphasis on the relativistic Hartree-Fock 
approximation) see \cite{thiesreview}. For a more recent review of static 
fermion bags in the GN model (with an emphasis on reflectionless backgrounds 
and supersymmetric quantum mechanics) see \cite{bagreview}. 
The large-$N$ semiclassical DHN spectrum of these fermion bags turns out to 
be essentially correct also for finite $N$, as analysis
of the exact factorizable S-matrix of the GN model reveals \cite{Smatrix}.

A variational calculation of these effects in the $1+1$ dimensional
massive generalization of the Gross-Neveu model, which we will
refer to as MGN, was carried in \cite{FZMGN} a few years ago, and more 
recently in \cite{Thies}. Very recently, we studied static fermion bags in 
the MGN model \cite{FH}, which we obtained using the inverse-scattering 
formalism, thus avoiding the need to choose a trial variational field 
configuration. Our main result in \cite{FH} is that the extremal static 
bag configurations are reflectionless, as in the massless Gross-Neveu model.
In the next section we briefly review the results of \cite{FH}, leaving 
technical details out. Then, in Section 3, we show that a subclass of the
reflectionless solitons of \cite{FH} arise self-consistently 
in the $1+1$ dimensional massive NJL (MNJL) model. The latter extends 
the results of \cite{shei, FZ-NJL} for the massless NJL model.
Solitons in the MNJL model were also recently studied in 
\cite{Thies}, where a derivative expansion was carried out around a particular
soliton background of the corresponding massless NJL model.

\section{Solitons in the Massive Gross-Neveu Model}

One way of writing the action for the MGN model is 
\beq
S =\int d^2x\,\left\{\sum_{a=1}^N\,\bar\psi_a\left[i\notpa-\si\right]\psi_a 
-{1\over 2g^2}\left(\si^2-2M\si\right)\right\}\,,
\label{lagrangian}
\eeq
where the $\psi_a$ ($a=1,\ldots,N$,) are massive Dirac fermions
and $\si$ is an auxiliary field. Integrating the $\si$ out results in an 
equivalent form of (\ref{lagrangian}), with quartic fermion self-interactions.

An obvious symmetry of (\ref{lagrangian}) with its $N$ Dirac spinors 
is $U(N)$. Actually, (\ref{lagrangian}) is symmetric under the larger group 
$O (2N)$ \cite{dhn} (see also Section 1 of \cite{bagreview}). 
The fact that the symmetry group of (\ref{lagrangian}) is $O (2N)$ rather than 
$ U(N)$ is related to the fact that it is invariant against 
charge-conjugation, like the massless GN model. Consequently, the energy 
eigenvalues of the Dirac equation associated with (\ref{lagrangian}), 
$[i\notpa-\si (x)]\,\psi = 0$, come in $\pm\om$ pairs. 

As usual, the theory (\ref{lagrangian}) can be rewritten with the help of 
the scalar flavor singlet auxiliary field $\si(x)$. Also as usual, we take
the large $N$ limit holding $\lambda\equiv Ng^2$ fixed. Integrating
out the fermions, we obtain the bare effective action
\beq
S[\si] =-{1\over 2g^2}\int\, d^2x
\,\left(\si^2-2M\si\right) -iN\,
\tr~{\rm log}\left(i\notpa-\si\right)\,.
\label{fermout}
\eeq
Noting that $\gam_5 ( i\notpa -\si )= -(i\notpa +\si)\gam_5
$,  we can rewrite the $\tr~{\rm log}(i\notpa -\si )$ as ${1\over 2}
\tr~{\rm log}\left[-(i\notpa -\si)(i\notpa +\si)\right]$. In this paper we 
focus on static soliton configurations. If $\si$ is time independent, the 
latter expression may be further simplified
to $ {T\over 2}\int {d\om \over 2 \pi} [\tr~{\rm log} (h_+-
\om^2)+\tr~{\rm log} (h_- -\om^2)]$ where 
$h_{\pm}  \equiv -\pa_x^2 + \si^2 \pm \si'\,,$
and where $T$ is an infra-red temporal regulator. 
As it turns out, the two Schr\"odinger operators $h_{\pm}$ are
isospectral (see Appendix A of \cite{bagreview} and Section 2 of 
\cite{josh1}) and thus we obtain
\beq
S[\si] = -{1\over 2g^2}\int\, d^2x
\,\left(\si^2-2M\si\right) - iNT\,
\int\limits_{-\infty}^{\infty} {d\om \over 2 \pi}\tr~{\rm log}(h_- -\om^2)\,.
\label{effective}
\eeq

In contrast to the standard massless GN model,
the MGN model studied here is not invariant under the $Z_2$
symmetry $\psi\rightarrow \gamma_5 \psi$, $\sigma \rightarrow -\sigma$,
and the physics is correspondingly quite different. As a result of the 
$Z_2$ degeneracy of its vacuum, the GN model contains 
a soliton (the so called CCGZ kink \cite{ccgz, dhn, others, josh1, bagreview})
in which the $\sigma$ field takes on equal and opposite values at 
$x=\pm\infty$. 
The stability of this soliton is obviously guaranteed by topological 
considerations. With any non-zero $M$ the vacuum value of $\sigma$ is 
unique and the CCGZ kink becomes infinitely massive and disappears. If any 
soliton exists at all, its stability has to depend on the energetics of 
trapping fermions.

Let us briefly recall the computation of the unique vacuum of the MGN model.
We shall follow \cite{FZMGN}. For an earlier analysis of the MGN ground state 
(as well as its thermodynamics), see \cite{klimenko}. Setting 
$\si$ to a constant we obtain from (\ref{effective}) the renormalized 
effective potential (per flavor)
$V(\si,\mu) = {\si^2\over 4\pi}~ {\rm log}~ {\si^2\over
e\mu^2} +
{1\over \lambda(\mu)}~\left[{\si^2\over 2} -
M(\mu)\si\right]\,,$
where $\mu$ is a sliding renormalization scale with
$\lambda(\mu)=Ng^2(\mu)$ and
$M(\mu)$ the running couplings. By equating the coefficient
of $\si^2$ in two versions of $V$, one defined with $\mu_1$ and the
other with $\mu_2$, we find immediately that
${1\over\lambda(\mu_1)} - {1\over\lambda(\mu_2)} =
{1\over \pi}~{\rm
log}
~{\mu_1\over\mu_2}$
and thus the coupling $\lambda$ is asymptotically free, just
as in the GN model. Furthermore, by equating the coefficient of $\sigma$
in $V$ we see that the ratio ${M(\mu)\over\lambda(\mu)}$
is a renormalization group invariant. Thus, $M$ and
$\lambda$ have the same scale dependence.

Without loss of generality we assume that $M(\mu)>0$  and
thus the absolute minimum of $V(\si,\mu)$, namely, the vacuum condensate
$m=\langle\si\rangle$, is the unique (and positive) solution of the gap 
equation
${dV\over d\si}~{\Big|_{\si=m}}= m\left[ {1\over \pi}~{\rm log}
~{m\over\mu} + {1\over \lambda(\mu)}\right] -
{M(\mu)\over\lambda(\mu)} = 0\,.$
Referring to (\ref{lagrangian}), we see that $m$ is the mass
of the fermion. Using the explicit scale dependence of $\lambda(\mu)$,
we can re-write the gap equation as
${m\over\lambda(m)}= {M(\mu)\over\lambda(\mu)}$,
which shows manifestly that $m$, an observable physical quantity, is a
renormalization group invariant. This equation also implies that $M(m)=m$, 
which makes sense physically.

Fermion bags correspond to inhomogeneous solutions of the saddle-point 
equation ${\delta S \over \delta \sigma(x,t)}=0$. In particular, static
bags $\sigx$ are the extremal configurations of the energy 
functional (per flavor) $\ce [\sigx] = - {S[\sigx]\over NT}\,,$
subjected to the boundary condition that $\sigx$ relaxes to its unique vacuum 
expectation value $m$ at $x=\pm\infty$. 
More specifically, we have to evaluate the energy functional of a static 
configuration $\sigx$, obeying the appropriate 
boundary conditions at spatial infinity, which supports $K$ pairs of bound 
states of the Dirac equation at energies $\pm\om_n$, $n=1,\ldots ,K$ 
(where, of course, $\om_n^2<m^2$). 
The bound states at $\pm\om_n$ are to be considered together, due to 
the charge conjugation invariance of the GN model. Due to Pauli's exclusion 
principle, we can populate each of the bound states $\pm\om_n$ with up to 
$N$ fermions. In such a typical multiparticle state, the 
negative frequency state is populated by $N-h_n$ fermions (i.e., by $h_n$ 
holes or antiparticles) and the positive frequency state contains $p_n$ 
fermions (or particles). We shall refer to the total number of particles and 
antiparticles trapped in the n-th pair of bound states 
$\nu_n = p_n+h_n $ as the valence, or occupation number of that pair.

The energy functional  $\ce [\sigx]$ is, in principle, a complicated and 
generally unknown functional of $\sigx$ and of its derivatives (which 
furthermore, requires regularization). Thus,   
the extremum condition ${\del\ce [\si]\over \del\sigx} =0$, as a functional
equation for $\sigx$, seems intractable. The considerable complexity of the 
functional equations that determine the extremal $\sigx$ configurations is 
the source of all difficulties that arise in any attempt to solve the model 
under consideration.  DHN found a way around this difficulty in the case of 
the GN model \cite{dhn}. They have used inverse scattering techniques 
\cite{inverse} to express the (regulated) energy functional 
$\ce [\si]$ in terms of the so-called 
``scattering data'' associated with, e.g., the hamiltonian $h_-$ mentioned 
above (and thus with $\sigx$), and then solved the extremum
condition ${\del\ce [\si]\over \del\sigx} =0$ with respect to those data.

The scattering data associated with $h_-$ are \cite{inverse} the 
reflection amplitude $r(k)$ of the Schr\"odinger operator $h_-$ at momentum 
$k$, the number $K$ of bound states in $h_-$ and their corresponding energies 
$0<\om_n^2\leq m^2\,,(n=1,\ldots, K)$, and also additional $K$ parameters 
$\{c_n\}$, where $c_n$ has to do with the normalization of the $n$th bound 
state wave function $\psi_n$ of $h_-$. More precisely, the $n$th bound 
state wave function, with energy $\om_n^2$, must decay as $\psi_n(x)\sim 
{\rm const.}\exp -\kappa_n x$ as $x\rightarrow\infty$, where  
$0<\kappa_n = \sqrt{m^2 -\om_n^2}\,.$
If we impose that $\psi_n (x)$ be normalized, this will determine the 
constant coefficient as $c_n$. (With no loss of generality, we may take 
$c_n>0$.) Recall that $r(-k) = r^*(k)$, since the Schr\"odinger potential 
$V(x) = \si^2(x) - \si'(x)$ is real. Thus, only the values of $r(k)$ for 
$k>0$ enter the scattering data. The scattering data are independent 
variables, which determine $V(x)$ uniquely, assuming $V(x)$ belongs to a 
certain class of potentials which fall-off fast enough toward infinity.
(Since the MGN does not bear topological solitons, neither $h_-$ nor $h_+$ 
can have a normalizable zero energy eigenstate. Thus, all the 
$\om_n$ are strictly positive.)

We can apply directly the results of DHN in order to write down that part
of $\ce [\sigx]$ which is common to the MGN and GN models, i.e., 
$\ce [\sigx]$ with its term proportional to $M$ removed, in terms of 
the scattering data. For lack of space we shall not write DHN's 
expression for the energy functional explicitly. Suffice it is to mention 
at this point that the ``DHN-part'' of $\ce [\sigx]$ depends on the 
reflection amplitude only via certain regular dispersion integrals of the 
quantity $\log [1-|r(k)|^2]$. The well-known reflectionless nature of 
solitons in the GN model is a direct consequence of this simple fact. 

In order to complete the task of expressing the effective action of the MGN 
model in terms of the scattering data, we have to find such a representation
for the remaining piece of $\ce [\sigx]$ proportional to $M$, 
namely, for $-{M\over\lambda}\int_{-\infty}^\infty \left(\sigx -m\right) 
\, dx$. The latter integral cannot be expressed in terms of the scattering 
data based on the trace identities of the Schr\"odinger operator 
$h_-$ discussed in Appendix B of \cite{dhn}. Evidently, new analysis is 
required to obtain its representation in terms of the scattering data. 
Happily enough, we were able to obtain such a representation in \cite{FH}, 
which reads
\beq\label{sigmatext}
\int\limits_{-\infty}^\infty \left(\sigx -m\right) \, dx = 
{1\over{2\pi i}}\int\limits_{-\infty}^\infty\,{\log \left[1-|r(k)|^2\right]
\over im-k}\,dk + 
\sum_{n=1}^K \log\left({m - \kappa_n \over m + \kappa_n}\right)\,.
\eeq
Thus, the $M$-dependent part of $\ce [\sigx]$, like its ``DHN-part'', 
depends on the reflection amplitude only via the combination 
$\log [1-|r(k)|^2]$. Combining these two terms together, it follows that 
$\delta \ce^{reg} [\sigx]/\delta r(k) = 
F(k)\, r^*(k)/(1-|r(k)|^2)$, where $F(k)$ is a calculable function, which 
does not vanish identically.
Thus, $r(k) \equiv 0$ is the unique solution of the variational equation 
$\delta \ce^{reg} [\sigx]/\delta r(k) = 0$. Static extremal bags in the 
MGN model are {\em reflectionless}, as their counterparts in the GN model.

Explicit formulas for reflectionless $\sigx$ configurations with an arbitrary
number $K$ of pairs of bound states are displayed in Appendix B of 
\cite{bagreview}. In particular, the one which supports a single pair of bound 
states at energies $\pm\om_b$ ($\kappa = \sqrt{m^2-\om_b^2}$), the one 
originally discovered by DHN, is
\beq\label{dhnsoliton}
\sigx = m + \kappa \left[ {\rm tanh }~\left( \kappa x -
{1\over 4}~{\rm log}~{m+\kappa\over m-\kappa}\right)- {\rm tanh }~
\left(\kappa x + {1\over 4}~{\rm log}
~{m+\kappa\over m-\kappa}\right)\right]\,.
\eeq

We see that the formidable problem of finding the extremal 
$\sigx$ configurations of the energy functional $\ce [\si]$ is reduced to the 
simpler problem of extremizing an ordinary function $\ce(\om_n, c_n) = 
\ce\left[\si (x; \om_n, c_n)\right]$ with respect to the 2$K$ parameters 
$\{c_n, \om_n\}$ that determine the reflectionless background $\sigx$. If we 
solve this ordinary extremum problem, we will be able to calculate the mass 
of the fermion bag. This we did in detail in \cite{FH}. Let us sketch the 
procedure and state the final result: 

The bare regulated energy 
function $\ce(\om_n)$ which depends on the bare couplings $\lambda$ and 
$M$ and on the UV-cutoff $\Lambda$ explicitly can be renormalized, in a manner
essentially similar to the effective potential, as was described above. 
$\ce $ is independent of the $c_n$'s, which appear in the scattering 
data. (The latter are thus flat directions for the energy function and 
determine the collective coordinates of the soliton.) The 
renormalized energy function thus obtained is a sum of the form $\sum_{n=1}^K
f(\om_n,\nu_n)$ where $f(\om,\nu)$ is a known function, which depends also
on the physical mass $m$ explicitly, and also through the RG-invariant ratio 
$\gamma = {1\over\lambda (m)} = {M(\mu)\over m\lambda(\mu)}$. 
Thus, the extremum condition fixes each $\om_n$ in terms of the number of 
the total number $\nu_n$ of particles and holes trapped in the bound states 
of the Dirac equation at $\pm\om_n$, and not by the numbers of trapped 
particles and holes separately (see (\ref{thetaext})). This fact is a 
manifestation of the 
underlying $O(2N)$ symmetry, which treats particles and holes symmetrically. 
Moreover, it indicates \cite{FH} that this pair of bound states gives rise to 
an $O(2N)$ antisymmetric tensor multiplet of rank $\nu_n$ of soliton states. 
(As it turns out, only tensors of ranks $0<\nu_n<N$ correspond to viable 
solitons \cite{FH}.) The soliton as a whole is therefore the tensor product 
of all these antisymmetric tensor multiplets. Finally, we showed in 
\cite{FH} that only the irreducible ($K=1$) soliton was protected by energy 
conservation and $O(2N)$ symmetry against decaying into lighter solitons 
(or free massive fermions). Its profile is given by (\ref{dhnsoliton}), where
$\kappa = m\sin \theta$ (or, equivalently, $\om = m\cos \theta$), with 
$0<\theta <\pi/2$, and where $\theta$ is determined by the extremum condition 
\beq\label{thetaext}
{\theta\over\pi} + \gamma\tan\theta = {\nu\over 2 N}\,.
\eeq
The left-hand side of (\ref{thetaext}) is a monotonically increasing function.
Therefore, (\ref{thetaext}) has a unique solution in the interval $[0,\pi/2]$. 
This solution is evidently smaller than $\theta^{\rm GN} = 
{\pi\nu\over 2N}$, the corresponding value of $\theta$ in the GN model
for the same occupation number. Thus, the corresponding bound state
energy $\om = m\cos\theta$ in the MGN model is higher than its
GN counterpart, and thus less bound. The soliton mass 
(i.e., the renormalized energy function, evaluated 
at the solution of (\ref{thetaext})) is 
\beq
\cm (\nu) = Nm
\left( {2\over \pi}\sin\theta + \gamma \log \,{1 + \sin\theta 
\over 1 - \sin\theta}\right)\,.
\label{solitonmass}
\eeq
This coincide with the corresponding 
results of variational calculations presented in \cite{FZMGN, Thies},
which were based on (\ref{dhnsoliton}) as a trial configuration. In fact, 
it was realized in \cite{Thies} that the trial configuration 
(\ref{dhnsoliton}) is an exact solution of the extremum condition 
${\del\ce [\si]\over \del\sigx} =0$, provided (\ref{thetaext})
is used to fix $\kappa = m\sin\theta $.

\section{Reflectionless solitons in the Massive Nambu-Jona-Lasinio Model}
It is natural to inquire whether the results of the previous section carry 
over to the phenomenologically interesting MNJL model. 
The action for the MNJL model may be written as a generalization of 
(\ref{lagrangian}),
$S =\int d^2x\,\left\{\sum_{a=1}^N\,\bar\psi_a\left[i\notpa-(\si + i\pi
\gamma_5)\right]\psi_a 
-{1\over 2g^2}\left(\si^2 +\pi^2-2M\si\right)\right\}\,,$
\label{lagrangian-njl}
where $\pix$ is a pseudo-scalar auxiliary field. (Here we assumed that the 
$2\times 2$ chiral mass matrix does not have a pseudo-scalar component, but
this does not restrict the generality of our discussion in any way. This 
particular orientation of the mass matrix can be always reached at by 
performing a global - and therefore, anomaly free- chiral rotation in the 
$\sigma-\pi$ plane.)

As in our discussion of the MGN model, we can integrate out the fermions,
and obtain the bare effective action 
\beq
S[\si] =-{1\over 2g^2}\int\, d^2x
\,\left(\si^2+\pi^2-2M\si\right) -iN\,
\tr~{\rm log}\left(i\notpa-\si- i\pi
\gamma_5\right)\,.
\label{fermout-njl}
\eeq
As before, we take the large $N$ limit, holding $\lambda\equiv Ng^2$ fixed. 
Unlike the NJL model, with its continuum of degenerate 
vacua, the ground state of the MNJL model (\ref{fermout-njl}) is unique, as 
in the MGN model. It corresponds to a constant field configuration, where 
$\pi=0$ and where $\si = m$ is determined by an equation identical to the one 
which arises in the MGN model.

Shei \cite{shei} has studied static solitons in the NJL model (i.e., 
$M=0$ in (\ref{fermout-njl})) using inverse scattering techniques. Similarly 
to DHN's results for the GN model, he concluded that extremal soliton 
profiles are reflectionless. Some of his results were rederived in 
\cite{FZ-NJL}, using a certain method based on properties of the diagonal 
resolvent of the Dirac operator, (which was applied first to the GN model in 
\cite{josh1}). 

Could Shei's analysis be extended to study solitons in the MNJL model, 
similarly to the extension of DHN's inverse scattering analysis to the MGN 
model? Could it be that the self-consistent static soliton backgrounds in the 
MNJL model are reflectionless? It seems that all we need in order to answer 
these questions is a generalization of (\ref{sigmatext}) to the case in which 
the Dirac operator involves a pseudo-scalar background $\pix$. 
Unfortunately, we were not able (so-far) to find such a generalization, and 
therefore we cannot answer these questions in general at the moment. 
However, we were able to find a particular family of self-consistent 
reflectionless static solitons in the MNJL model by making an educated guess, 
as we shall now explain.

The spectrum of the Dirac equation associated with (\ref{fermout-njl}) is 
{\em not} invariant against charge-conjugation, unless $\pix\equiv 0$. 
Thus, the bound states corresponding to a static soliton background are not 
paired, in general. In particular, as has been shown by Shei, there exist 
solitons in the NJL model which bind fermions into a single bound state. 
However, he has also found solitons with charge-conjugation-invariant 
spectrum (see Eqs.(3.25)-(3.28) in 
\cite{shei}), with a pair of bound states $\pm\om_b$, in which $\pix=0$ 
identically, and $\sigx$ is given by (\ref{dhnsoliton}), which thus coincide 
with the DHN solitons in the GN model, for which $\om_b= 
m \cos\left({\pi\nu\over 2N}\right)$ and $\cm_{\rm DHN} (\nu) =  
{2 Nm\over \pi}\sin\left({\pi\nu\over 2N}\right)$. However, unlike in the 
GN model, in the NJL model we must choose $p=h={\nu\over 2}$ (i.e., a soliton
of this type must trap an equal number of fermions and anti-fermions).
The reason for this restriction is not hard to understand physically: 
the total chiral rotation $\Delta\theta$, namely, the difference in 
$\arctan {\pix\over\sigx}$ between the two ends of the one dimensional space, 
must be related to the fermion number charge $n_f$ deposited in the soliton
according to $\Delta\theta = -{2\pi n_f\over N}$ \cite{goldwil} (see also Eqs.
(5.10) and (5.22) in \cite{FZ-NJL}). 
The soliton profile $(\sigx,\pix)$ under consideration, starts at the vacuum 
point $(m,0)$ at $x=-\infty$ and returns to it at $x=+\infty$. Thus, 
$\Delta\theta = n_f = p-h = 0$ for this soliton.

Now, any static soliton profile in the MNJL model must start at the 
{\em unique} vacuum  $(m,0)$ at $x=-\infty$ and return to it at $x=+\infty$. 
Thus, it should bring about null total chiral rotation, precisely as Shei's
charge-conjugation-invariant configuration does. Therefore, if the MNJL bears
{\em reflectionless} static solitons, they must be of this form (or 
charge-conjugate invariant generalizations thereof, with more pairs of 
paired bound states). The only thing that should change compared to the 
NJL model is the quantization condition, relating $\om_b$ and $\nu$.

We have verified that this is indeed the case, simply by substituting this 
configuration into the static inhomogeneous saddle-point equations associated 
with (\ref{fermout-njl}). Varying (\ref{fermout-njl}) with respect to 
$\pix$ we {\em obviously} obtain an equation identical in form to that of the 
NJL model. (For the latter, see the second equation in 
(5.1) in \cite{FZ-NJL}). Using the explicit expressions for the entries of 
the diagonal resolvent of the Dirac operator with a reflectionless 
$(\sigx, \pix)$ background with two bound states (Eqs.
(4.13) and (4.14) in  \cite{FZ-NJL} with paired bound states 
$\om_2 = -\om_1 $), we see that $\pix\equiv 0$ is indeed a solution of that
equation.
(Here, having $p=h={\nu\over 2}$ is essential.). This $\pi$-equation leaves
$\om_1$ an undetermined function of $\nu$. We still have to vary with respect
to $\sigx$. Substituting the explicit expressions for the appropriate 
entries of the diagonal resolvent of the Dirac operator (Eqs.(4.13), (4.14) 
and (2.10) of \cite{FZ-NJL}) in the saddle-point equation arising from 
variation with respect to $\sigx$, and using the 
simplifying identity Eq. (2.24) of \cite{jmp}, we arrive simply at the 
static saddle-point equation of the MGN model, which is solved by $\sigx$ 
given by (\ref{dhnsoliton}) and the quantization condition (\ref{thetaext}),
leading to soliton mass (\ref{solitonmass}). Thus, a restricted subset
of the extremal reflectionless solitons of the MGN model appear, 
not surprisingly, also in the MNJL model. For these solitons $\pix\equiv 0$. 
The question whether these solitons exhaust all possibilities in the 
two-dimensional MNJL model remains open. 


\end{document}